\documentstyle[12pt,epsfig]{article}
\newcounter{subequation}[equation]

\makeatletter 
 
\def\thesubequation{\theequation\@alph\c@subequation}
\def\@subeqnnum{{\rm (\thesubequation)}}
\def\slabel#1{\@bsphack\if@filesw {\let\thepage\relax
   \xdef\@gtempa{\write\@auxout{\string
      \newlabel{#1}{{\thesubequation}{\thepage}}}}}\@gtempa
   \if@nobreak \ifvmode\nobreak\fi\fi\fi\@esphack}
\def\subeqnarray{\stepcounter{equation}
\let\@currentlabel=\theequation\global\c@subequation\@ne
\global\@eqnswtrue
\global\@eqcnt\z@\tabskip\@centering\let\\=\@subeqncr
$$\halign to \displaywidth\bgroup\@eqnsel\hskip\@centering
  $\displaystyle\tabskip\z@{##}$&\global\@eqcnt\@ne
  \hskip 2\arraycolsep \hfil${##}$\hfil
  &\global\@eqcnt\tw@ \hskip 2\arraycolsep
  $\displaystyle\tabskip\z@{##}$\hfil
   \tabskip\@centering&\llap{##}\tabskip\z@\cr}
\def\endsubeqnarray{\@@subeqncr\egroup
                     $$\global\@ignoretrue}
\def\@subeqncr{{\ifnum0=`}\fi\@ifstar{\global\@eqpen\@M
    \@ysubeqncr}{\global\@eqpen\interdisplaylinepenalty \@ysubeqncr}}
\def\@ysubeqncr{\@ifnextchar [{\@xsubeqncr}{\@xsubeqncr[\z@]}}
\def\@xsubeqncr[#1]{\ifnum0=`{\fi}\@@subeqncr
   \noalign{\penalty\@eqpen\vskip\jot\vskip #1\relax}}
\def\@@subeqncr{\let\@tempa\relax
    \ifcase\@eqcnt \def\@tempa{& & &}\or \def\@tempa{& &}
      \else \def\@tempa{&}\fi
     \@tempa \if@eqnsw\@subeqnnum\refstepcounter{subequation}\fi
     \global\@eqnswtrue\global\@eqcnt\z@\cr}
\let\@ssubeqncr=\@subeqncr
\@namedef{subeqnarray*}{\def\@subeqncr{\nonumber\@ssubeqncr}\subeqnarray}
\@namedef{endsubeqnarray*}{\global\advance\c@equation\m@ne%
                           \nonumber\endsubeqnarray}

\makeatletter
\@addtoreset{equation}{section}
\makeatother
\renewcommand{\theequation}{\thesection.\arabic{equation}}

\def\dalemb#1#2{{\vbox{\hrule height .#2pt
        \hbox{\vrule width.#2pt height#1pt \kern#1pt
                \vrule width.#2pt}
        \hrule height.#2pt}}}
\def\square{\mathord{\dalemb{6.8}{7}\hbox{\hskip1pt}}}

    \let\e=\epsilon
  \let\q=\theta  
  \let\n=\nu

 \def\bd{\begin{document}} \def\ed{\end{document}}
\def\ds{\documentstyle} \let\fr=\frac \let\bl=\bigl \let\br=\bigr
\let\Br=\Bigr \let\Bl=\Bigl 
\let\bm=\bibitem
\let\na=\nabla
\let\pa=\partial \let\ov=\overline
\def\ie{{\it i.e.\ }} 
\newcommand{\be}{\begin{equation}} 
\newcommand{\ee}{\end{equation}} 
\def\ba{\begin{array}}
\def\ea{\end{array}}
\def\ft#1#2{{\textstyle{{\scriptstyle #1}\over {\scriptstyle #2}}}}
\def\fft#1#2{{#1 \over #2}}
\def\del{\partial}
\def\sst#1{{\scriptscriptstyle #1}}
\def\oneone{\rlap 1\mkern4mu{\rm l}}
\def\e7{E_{7(+7)}}
\def\td{\tilde}
\def\wtd{\widetilde}
\def\im{{\rm i}}
\def\bog{Bogomol'nyi\ }
\def\q{{\tilde q}}
\def\hast{{\hat\ast}}
\def\0{{\sst{(0)}}}
\def\1{{\sst{(1)}}}
\def\2{{\sst{(2)}}}
\def\3{{\sst{(3)}}}
\def\4{{\sst{(4)}}}
\def\5{{\sst{(5)}}}
\def\6{{\sst{(6)}}}
\def\7{{\sst{(7)}}}
\def\8{{\sst{(8)}}}
\def\n{{\sst{(n)}}}
\def\oo{{\"o}}
\def\hA{\hat{\cal A}}
\def\ns{{\sst {\rm NS}}}
\def\rr{{\sst {\rm RR}}}
\def\tH{{\widetilde H}}
\def\tB{{\widetilde B}}
\def\cA{{\cal A}}
\def\cF{{\cal F}}
\def\tF{{\wtd F}}
\def\Z{\rlap{\sf Z}\mkern3mu{\sf Z}}
\def\ep{{\epsilon}}
\def\IIA{{\rm IIA}}
\def\IIB{{\rm IIB}}
\def\ads{{\rm AdS}}
\def\R{\rlap{\rm I}\mkern3mu{\rm R}}
\def\mapright#1{\smash{\mathop{-\!\!\!-\!\!\!-\!\!\!-\!\!\!-\!\!\!
             \longrightarrow}\limits^{#1}}}

\def\Ei{{\hbox{Ei}}}
\def\Ci{{\hbox{Ci}}}
\def\Si{{\hbox{Si}}}

\newcommand{\ho}[1]{$\, ^{#1}$}
\newcommand{\hoch}[1]{$\, ^{#1}$}
\newcommand{\bea}{\begin{eqnarray}} 
\newcommand{\eea}{\end{eqnarray}} 
\newcommand{\ra}{\rightarrow}
\newcommand{\lra}{\longrightarrow}
\newcommand{\Lra}{\Leftrightarrow}
\newcommand{\aap}{\alpha^\prime}
\newcommand{\bp}{\tilde \beta^\prime}
\newcommand{\tr}{{\rm tr} }
\newcommand{\Tr}{{\rm Tr} } 
\newcommand{\NP}{Nucl. Phys. }
\newcommand{\tamphys}{\it Center for Theoretical Physics,
Texas A\&M University, College Station, TX 77843}

\newcommand{\brussels}{\it Physique Th\'eorique et Math\'ematique, 
Universit\'e Libre de Bruxelles,\\ Campus Plaine C.P. 231, B-1050
Bruxelles, Belgium} 

\newcommand{\auth}{J. F. V\'azquez-Poritz}

\begin{document}
\begin{flushright}
ULB-TH/01-34\\
November  2001\\
\hfill{\bf hep-th/0110299}\\
\end{flushright}


\begin{center}

{\large {\bf Massive Gravity on a Non-extremal Brane}}

\vspace{20pt}

\auth

\vspace{10pt}
\brussels\\
\vspace{10pt}

\vspace{30pt}

\underline{ABSTRACT}

\end{center}

We consider a brane world scenario which arises as the near-horizon region
of a non-extremal D5-brane. There is a quasi-localized massive
graviton mode, as well as harmonic modes of higher mass which are bound to
the brane to a lesser degree. Lorentz invariance is slightly broken, which
may have observable effects due to the leakage of the metastable graviton
states into the bulk. Unlike for a brane world arising from an extremal
D5-brane, there is no mass gap. We also find that a brane world arising
from a non-extremal M5/M5-brane  intersection has the same graviton
dynamics as that of a non-extremal D5-brane. This is evidence that a
previously conjectured duality relation between the dual quantum field
theories of each p-brane background may hold away from extremality. 

{\vfill\leftline{}\vfill
\vskip 10pt \footnoterule {\footnotesize \hoch{1} This work is supported
in part by Francqui Foundation (Belgium), the Actions de Recherche
Concert{\'e}es of the Direction de la Recherche Scientifique - Communaut\'e
Francaise de Belgique, IISN-Belgium (convention 4.4505.86).
 
\vskip  -12pt} \vskip   14pt
}

\pagebreak
\setcounter{page}{1}


\section{Introduction}

Randall and Sundrum \cite{randall1,randall2} have shown that, with fine
tuned brane tension, a flat 3-brane embedded in $AdS_5$ can have a single
massless bound state. Four-dimensional gravity is recovered at low-energy
scales. It has also been proposed that part or all of gravitational
interactions are the result of massive gravitons. For example, in one
model, gravitational interactions are due to the net effect of the
massless graviton and ultra-light Kaluza-Klein state(s) 
\cite{kogan1,mous,kogan2,kogan3}. In another proposal, there is
no normalizable massless graviton and four-dimensional gravity is
reproduced at intermediate scales from a resonance-like behavior of the
Kaluza-Klein wave functions \cite{kogan2,kogan3,gregory,csaki2,dvali}.

It has been shown that an $AdS_4$ brane in $AdS_5$ does not have
a normalizable massless graviton. Instead, there is a very light, but
massive bound state graviton mode, which reproduces four-dimensional
gravity \cite{karch,kogan,kogan4,porrati1}. The bound state mass as a
function of brane tension, as well as the modified Newtonian
gravitational potential, were explored in \cite{schwartz,miemiec}.

Five-dimensional domain walls which localize gravity may arise from a sphere
reduction from ten or eleven dimensions, as the near-horizon of extremal
$p$-branes \cite{cvetic}. If such a $p$-brane is perturbed away from
extremality, what effects would this have on the corresponding brane world?
In this paper, we consider a brane  world scenario which arises
from a sphere reduction of the near-horizon region of a non-extremal
D5-brane. This is a new example in which there is a massive bound graviton
state. There is also the novel phenomenon of harmonic graviton modes of
higher mass, which are bound to the brane to a lesser degree.

It has already been proposed that our observable world resides within the
horizon of a non-extremal brane, in order to account for the small but
nonzero entropy density of our universe \cite{moon1,moon2}. Hawking
radiation would cause the brane to evolve into an extremal state, which
could be a possible mechanism for the resolution of various problems
associated with brane world scenarios, such as the observed flatness and
approximate Lorentz invariance of our world. While our particular model
embeds the brane world outside of the D5-brane horizon, much of the
motivations of \cite{moon1,moon2} remain the same.

This paper is organized as follows. In section 2, we discuss how a
deviation from extremality breaks Lorentz invariance, which may have
observable effects due to the leakage of the (almost) massless graviton mode
into the bulk \cite{rubakov}. In section 3, we show that there is a
massive graviton quasi-localized on the brane world, as well as harmonic
modes of higher mass that are bound to the brane to a lesser
degree. Sections 2 and 3 focus on the brane world arising from the
near-horizon region of a non-extremal D5-brane wrapped on $T^2$. In
section 4, we consider a non-extremal M5/M5-brane intersection wrapped on
$T^3$ or $K3$. In section 5, we have concluding remarks. 

\section{Non-extremality and violation of Lorentz invariance}

The metric for a non-extremal D5-brane is given by
\be
ds^2=H^{-1/4}(-f dt^2+dx_i^2)+H^{3/4}(f^{-1} dr^2+r^2 d\Omega_3^2),
\label{metric}
\ee
where 
\be
H=1+\frac{R^2}{r^2},\ \ \ \ \ \ \ \ \ f=1-\frac{e^2 R^2}{r^2},
\ee
and $i=1,..,5$. $e$ is the non-extremality parameter. For $e \ll 1$, we
can neglect the "1" in $H$ in the near-horizon limit. If we make the
coordinate transformation
\be
\frac{r}{R}=\sqrt{{\rm e}^{-k|z|}+e^2}, \label{coord1}
\ee
then the metric (\ref{metric}) in the near-horizon region can be written
as
\be
ds^2=({\rm
e}^{-k|z|}+e^2)^{1/4}[f(-dt^2+dz^2)+dx_i^2+R^2d\Omega_3^2]. \label{metric2}
\ee
In the extremal limit $e=0$, the metric (\ref{metric2}) is expressed in
the conformally-flat frame. $z=0$ corresponds to $r/R=\sqrt{1+e^2}$, which
is the location of the brane-world. $z \rightarrow \infty$ corresponds to
$r \rightarrow e R$, which was the horizon of the D5-brane before the
near-horizon limit was taken. We will consider $x_4$ and $x_5$ to be
wrapped around a compact manifold, so that the inhabitants of this brane
world would observe their universe to be $1+3$-dimensional. That is, the
D5-brane is dimensionally reduced on $T^2 \times S^3$.

Note that the metric in (\ref{metric2}) does not exhibit
four-dimensional Lorentz invariance in the brane world
directions. However, the four-dimensional geometry on the brane world is
(approximately) invariant under Lorentz symmetry, since at the brane world
position $z=0$ the factor in front of the $dt^2$ becomes
$f(z=0)=1/(1+e^2)$, which can readily be absorbed into the time coordinate
\cite{rubakov}. 

The Laplacian on the non-compact world-volume coordinates is
\be
\square_4=f^{-1}\partial_t^2 -\partial_{x_1}^2 -\partial_{x_2}^2
-\partial_{x_3}^2=f^{-1}E^2-\vec{p}\ ^2.
\ee
The non-extremality breaks the bulk Lorentz invariance in such a way that
the momentum term can be neglected at large $z$ relative to the energy
term. Thus, unlike the extremal limit in which there is a mass gap
\cite{cvetic}, the mass spectrum is continuous from $m=0$. Furthermore,
the massless graviton is no longer bound to the brane forever. This mode 
is quasi-localized for nonzero momentum. Such leakage into the extra
dimension implies that the four-dimensional Lorentz invariance is only
approximate \cite{rubakov,dub}. 

Another Lorentz-violating effect is the modification of the dispersion
relation to
\be
E^2=m^2+c^2 \vec{p} ^2,
\ee
where $c$ depends on the spread of the wave function in the extra
dimension \cite{rubakov,csaki,cole}, and $c=1$ in the extremal case. For a
prototype model of a scalar field $\phi_m (z)$ considered in \cite{rubakov},
our non-extremal geometry yields the result
\be
c^2=1-\frac{e^2 k}{(1+e^2)^{3/2}} \frac{\int dz z |\phi_m (z)|^2}{\int dz
|\phi_m (z)|^2},
\ee
in the approximation of a narrow wave function.

\section{Localization of massive graviton}

The equation of motion for a graviton fluctuation is
\be
\partial_M \sqrt{-g}g^{MN}\partial_N \Phi=0, 
\ee
where the wave function $\Phi$ depends on the non-compact coordinates
$t, z, x_1, x_2, x_3$. We take $\Phi=\phi(z)M(t,x_1,x_2,x_3)$, where
$\square_{(4)} M=m^2 M$ and $\square_4$ is the Laplacian on
$t,x_1,x_2,x_3$.

For the background (\ref{metric2}) the wave equation is
\be
-{\rm e}^{k|z|}\partial_z({\rm e}^{-k|z|}+e^2)\partial_z\phi=m^2
\phi. \label{wave}
\ee
With the wave function transformation
\be
\phi=({\rm e}^{-k|z|}+e^2)^{-1/2}\psi,
\ee
the wave equation (\ref{wave}) becomes
\be
-\partial_z^2 \psi+ \psi+V(z) \psi=0, 
\ee
where the effective potential
\be
V(z)=\Big[\frac{k^2/2-m^2}{1+e^2 {\rm e}^{k|z|}}-\frac{k^2}{4(1+e^2 {\rm
e}^{k|z|})^2}-\frac{k}{1+e^2}\delta (z) \Big], \label{V}
\ee
which includes the mass term. For the massless mode, this is a
Schr\"{o}dinger-type wave equation, with the solution
\be
\psi=N(e^2+{\rm e}^{-k|z|})^{1/2}.
\ee
For the extremal case, this wave function is normalizable, with $N$ being
the normalization constant. The massless graviton is
therefore bound to the extremal brane world. 
\begin{figure}
   \epsfxsize=4.0in
   \centerline{\epsffile{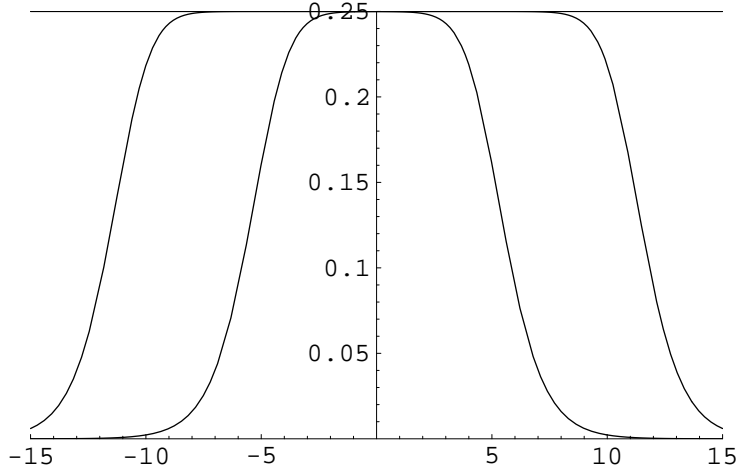}}
   \caption[FIG. \arabic{figure}.]{$V(z)$ for $m/k=0$ and $e=0,.005,.01$}
\end{figure}
\begin{figure}
   \epsfxsize=4.0in
   \centerline{\epsffile{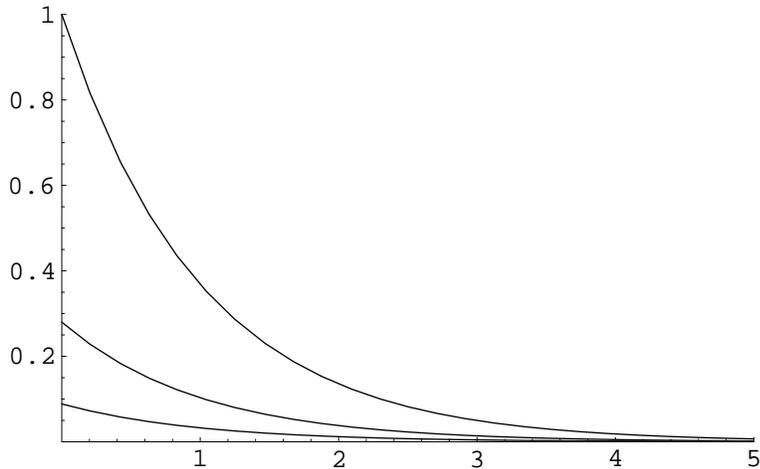}}
   \caption[FIG. \arabic{figure}.]{$|\psi (z)|^2$ for $m/k=0$ and
$e=0,.005,.01$}
\end{figure}
The corresponding effective potential (\ref{V}) is constant (for
$e=0$) with respect to $z$, as shown in Figure 1. 

For non-extremality, the effective potential
decreases to zero as $|z| \rightarrow \infty$, as is depicted in Figure 1
for the cases $e=.005$ and $.01$. Away from extremality, $\psi$ is not
normalizable and the massless graviton propagates in the extra dimension
$z$. This can be seen from Figure 2, in which the relative peak of $|\psi
|^2$ at $z=0$ is diminished away from extremality.
\begin{figure}
   \epsfxsize=4.0in
   \centerline{\epsffile{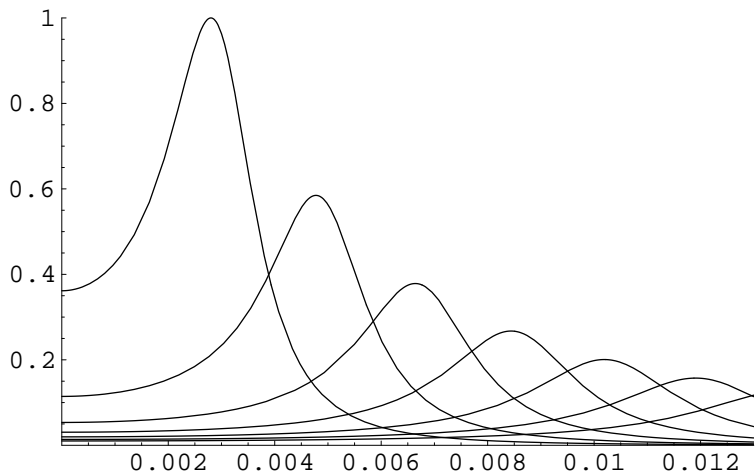}}
   \caption[FIG. \arabic{figure}.]{$|\psi (z=0)|^2$ versus $m/k$ for
$e=.005,.01,.015,.02,.025,.03,.035$}
\end{figure}

For the massive gravitons, the wave equation cannot be solved analytically
and so we content ourselves with plotting numerical solutions. We
numerically solve the wave equation outwards from $z=0$ with the boundary
conditions $\psi(z=0)=1$\\ ($\psi$ is not yet normalized) and 
$\partial_z \psi (z=0)=-k/(2+2e^2)$, the second of which is due to the
$\delta(z)$. We then numerically integrate in order to find the correct
normalization factor.
\begin{figure}
   \epsfxsize=4.0in
   \centerline{\epsffile{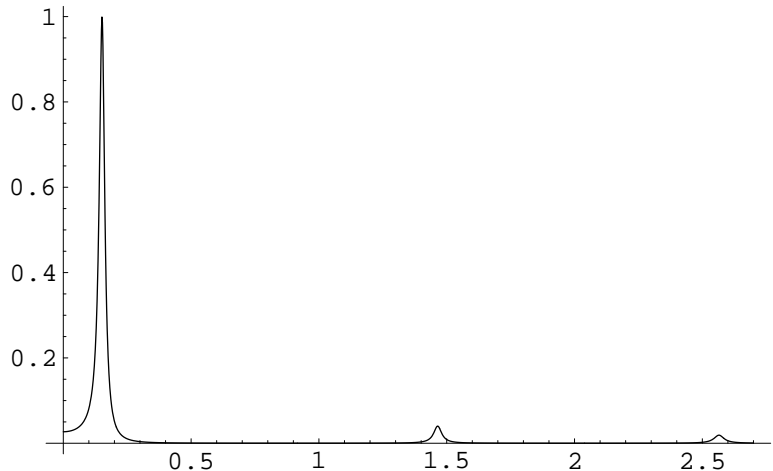}}
   \caption[FIG. \arabic{figure}.]{$|\psi (z=0)|^2$ versus $m/k$ for $e=.5$}
\end{figure}
As can be seen in Figure 3, away from extremality there is a resonance in
the wave function at nonzero mass, which represents a massive graviton
mode localized on the brane world. The resonance mass increases as one moves
away from extremality. Also, moving away from extremality has the effect
of dissipating the wave function, which implies a decrease in the distance
that the massive graviton propagates on the brane world before escaping into
the extra dimension.

As the non-extremality parameter $e$ increases, harmonic graviton states
arise which are quasi-localized on the brane world, as is shown in Figure
4. However, for higher bound harmonics, the corresponding distance which
the graviton mode travels on the brane world before escaping is
diminished. The presence of these harmonic bound states is a novel feature
in brane world scenarios. 

\section{D5-brane versus M5/M5-brane intersection}

Supergravity domain-wall solutions with exponential scalar potentials
can sometimes have the higher-dimensional origin as the near-horizon region
of a $p$-brane configuration. It has been found that the near-horizon
region of the extremal D5-brane reduced on $T^2 \times S^3$ yields the same
exponential scalar potential as that of the extremal M5/M5-brane
intersection reduced on $T^4 \times S^2$ or $K3 \times S^2$. This may hint
that there is a duality relation between the quantum field theory living on
the noncompact world volume of these two brane configurations
\cite{cvetic}.

We investigate this issue further by considering the non-extremal
generalizations of the above $p$-brane configurations, from the point of
view of the localization of gravity. Much of what has been written
previously about the brane world arising from a non-extremal D5-brane also
applies to the case of the non-extremal M5/M5-brane intersection. The
metric for the latter is
\bea
ds^2&=&H_1^{-1/3}H_2^{-1/3}(-f dt^2+dx_i^2)+H_1^{-1/3}H_2^{2/3}dy_j^2+\\
\nonumber & & 
H_1^{2/3}H_2^{-1/3}dw_j^2+H_1^{2/3}H_2^{2/3}(f^{-1}dr^2+r^2d\Omega_2^2),
\eea
where 
\be
H_i=1+\frac{R_i}{r},\ \ \ \ \ \ \ \ \ f=1-\frac{g R}{r},
\ee
and $i=1,2,3$ and $j=1,2$. $g$ is the non-extremality parameter. In the
near-horizon limit, we can neglect the "1" in $H_i$ and make the coordinate
transformation
\be
\frac{r}{R}={\rm e}^{-k|z|}+g, \label{coord2}
\ee
which is analogous to (\ref{coord1}) for the D5-brane. The corresponding
wave equation for a graviton fluctuation is 
\be
-{\rm e}^{k|z|}\partial_z({\rm e}^{-k|z|}+g)\partial_z\phi=m^2
\phi.
\ee
If we take $g \rightarrow e^2$, then the above wave equation is identical
to that for the D5-brane given by (\ref{wave}). Thus, from the point of
view of localization of gravity, a duality relation between the quantum
field theory living on the world volume of the D5-brane wrapped on $T^2$
and that of the $M5/M5$-brane intersection wrapped on $T^4$ or $K3$ may
hold away from extremality, provided that $g$ is identified with $e^2$.

\section{Conclusions}

We have considered a brane world scenario which arises from the
near-horizon region of a non-extremal D5-brane wrapped around
$T^2$. Unlike for the extremal D5-brane, the graviton spectrum has no mass
gap. There is a quasi-localized massive graviton mode, whose mass increases
continuously from zero as one moves away from extremality. In addition,
there are harmonic graviton modes which are bound to the brane to a lesser
degree. In this scenario, Lorentz invariance is slightly broken, effects
of which may be communicated from the bulk to the brane via the leakage of
metastable graviton states. It would be interesting to see if the same
graviton dynamics occur for a brane world arising from other $p$-brane
origins, such as the near-horizon region of a D4-brane wrapped around 
$S^1$ or that of a D3-brane. However, in most cases the dynamical equations
in non-extremal $p$-brane backgrounds are difficult to handle. 

However, we do find the same graviton dynamics happening for a brane world
arising from the near-horizon region of a non-extremal M5/M5-brane
intersection wrapped around $T^4$ (or $K3$). In fact, this provides
evidence that a conjectured duality relation between the dual quantum field
theories living on the world volume of a D5-brane wrapped around $T^2$ and
an M5/M5-brane intersection wrapped around $T^4$ may hold away from
extremality.

This duality was initially conjectured because the dimensional reductions
of the near-horizon regions of these two $p$-brane configurations down to
five dimensions yield the same exponential scalar potential
\cite{cvetic}. Such duality relations between various $p$-branes and
$p$-brane intersections may exist for other cases. This is certainly worthy
of further investigation, since there may be a more general class of
examples for which a given gravity solution/QFT pair may be dual to one or
more such pairs.

\section*{Acknowledgments}

I would like to thank M. Henneaux for proof-reading this paper, and
M. Cveti\v{c} for a useful discussion. I am also grateful to V.A. Rubakov
for a very helpful discussion at Conference Francqui 2001 in Bruxelles,
Belgium.

\end{document}